# A Canonical Representation of Data-Linear Visualization Algorithms

Thomas Baudel, *IBM*

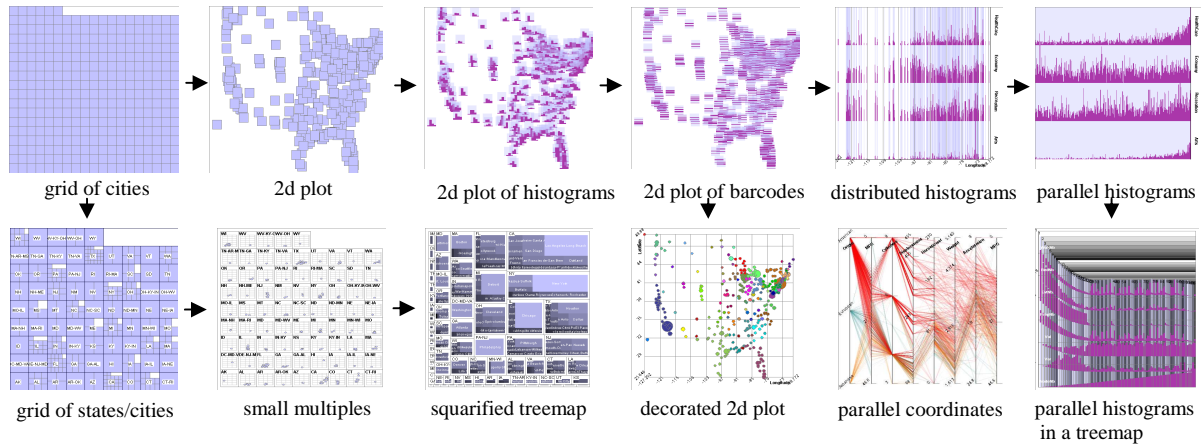

Fig. 1. Chains of visualization transformations. Switching from one view to the next requires changing 4 parameters at most.

**Abstract**— We introduce *linear-state dataflows*, a canonical model for a large set of visualization algorithms that we call *data-linear visualizations*. Our model defines a fixed dataflow architecture: partitioning and subpartitioning of input data, ordering, graphic primitives, and graphic attributes generation. Local variables and accumulators are specific concepts that extend the expressiveness of the dataflow to support features of visualization algorithms that require state handling. We first show the flexibility of our model: it enables the declarative construction of many common algorithms with just a few mappings. Furthermore, the model enables easy mixing of visual mappings, such as creating treemaps of histograms and 2D plots, plots of histograms… Finally, we introduce our model in a more formal way and present some of its important properties. We have implemented this model in a visualization framework built around the concept of linear-state dataflows.

——————————— ✦ ———————————

## 1 INTRODUCTION

An important research challenge in information visualization is the production of descriptive models to explore and compare visualizations in a design space of limited dimensionality. Indeed, in the wide range of existing visualization methods, many may be equivalent; few are actually supported by refutable observations, and even fewer have been formally evaluated in a rigorous context. The ability to position visualizations in a well-defined space should help assessing their relative merits according to rational criteria.

We are also driven by more pragmatic goals. One of the essential features of information visualization systems is flexibility. Many existing systems allow interactive adjustments of visualization parameters to match the information display to perceptual capabilities and task requirements. We aim at providing concise means to describe all possible visualizations of certain classes in a declarative manner.

Various directions to address these issues have been explored, relying on graphical grammars [5, 35], relational algebra [20, 30, 31] or a software architecture reference model [7, 11, 15].

We introduce a canonical, computational model for the description of a large set of information visualization algorithms. We call this set *data-linear visualizations* and our model *linear-state dataflows*.

Our approach echoes the proposition of relational calculus (SQL's formalization) for database querying: rather than relying on a Turing-complete model, the relational algebra offers only 3 operators which can be combined to declaratively produce a wide range of queries, deemed sufficient for many common purposes. Reliance on a soundly defined computational model allows precise assessment of the power of expression of the resulting algebra, in contrast to inconsistency and lack of predictability resulting from proposing an extensible set of ad-hoc operators.

First we present our research background: dataflow architectures and flexible interactive visualization systems. Then, we introduce our model with a walkthrough, showing how to describe many widely used visualizations in a very concise way, as well as how to cross features from various types of visualizations, such as creating treemaps [27] of 2D graphs, grids of parallel coordinate systems or histograms of 2D plots. Finally, we propose some definitions and present some core properties of our model: canonicity and completeness for an important set of visualization algorithms.

- *Thomas Baudel is Researcher at IBM, Software Group, ILOG products.*
  *E-mail: baudelth@fr.ibm.com*

## 2 BACKGROUND

### 2.1 Dataflows in Visualization

Dataflows to provide conciseness and flexibility in the construction of representations is as old as scientific visualization: IDL [23] and PV-Wave [33], whose designs date back to the 1980's, are still widely used tools. On newer platforms, toolkits such as Protovis [7], Axiis [14], and many others, provide similar dataflow architectures directly in mainstream development languages. These languages and toolkits provide the user with data set operators that can be connected to one another and transform or organize the data sets in intermediary representations up to a final visual display. These operators follow a classical reference model [9]: data transforms, primitive generation, decoration of primitives, view transform and rendering. Some of these toolkits rely on a well defined grammar of operators, such as VizQL [30], which is based on the relational algebra.

Interactive applications on top of these dataflow architectures, such as VisAd [16], VTK [26], Tioga-2 [2], Visage [24], and many others, have added the ability to build visualizations interactively. These tools offer scripting languages and have allowed the design of "Visualization Spreadsheets" [10] which allow users to interactively assemble filter and rendering operations. Creating visualizations with these tools is still complex, as the capabilities of the tools are unbounded: operators can be combined almost arbitrarily and the language can be extended with custom operators providing a lot of power, but preventing, in our opinion, the user's creation of a closed mental model of the tool's capabilities.

The model we propose, which we call linear-state dataflows, offers only *some* of the capabilities found in the above mentioned tools. It allows partitioning, accumulation of values, sorting data sets and assigning, conditionally or not, graphic primitives and decorations, in a recursive fashion. Besides these basic operations, we introduce the notion of local variables and accumulators, which enable the description of some visual features that require state handling. By design, these can only be used to implement graphic features that depend on the input data in a linear fashion. Hence, our model can only describe visualizations rendered in at most a linear function of the size of the input data. The converse proposition is true: any data-linear visualization algorithm can be described in our model, as we will see later.

### 2.2 Interactive Editing of Visualizations

Interactive manipulation of display parameters is a base requirement in information visualization. As J. Bertin puts it, as early as 1969: "Modern graphics is about *transformable* and reorderable images [...] The computer realizes there its most complete and powerful expression" [5].

Since SPAD Amado [12, 21], many tools have provided the ability to define dynamic and flexible representations of data sets: Spotfire [29], Advizor [1, 13], xgobi [25], Table Lens [20], the Attribute Explorer, Influence Explorer and Prosection Matrix [28], Improvise [34] or Polaris & Tableau [31]. Each of these tools provides visualization methods whose parameters (coordinates, color…) can be mapped to individual data dimensions or aggregates.

Our model allows reproducing most of the above mentioned visualization methods through parameter settings. It also enables creating mixed visualizations that combine features of different methods, such as a treemap embedding an image file or a matrix of 2D graphs. Many of these representations are certainly quite exotic; nevertheless they can be obtained through the setting of a fixed and ordered number of parameters. Hence, we may suppose that a user could gain a reasonable conceptual model of the tool by working on the effect and interplays of each parameter one by one. Using the interactive implementation of our model [17], advanced users can browse the space of possible representations of a data set (Figure 1), or create and embed visualizations in an external application, providing final users with a restricted subset of customizations.

The parameters of our rendering model are expressions that can include attributes of the data set and local variables. In this sense, users do not customize one particular algorithm, but describe the main features of an algorithm. The fixed set of parameters defines a class of visualizations that can be defined in our model. This approach is declarative as one needs not specifying control flow or dataflow connections, but only essential characteristics of a program.

## 3 SOME DEFINITIONS

### 3.1 Data models and Data tables

Our model takes as input a data table. The user provides a homogeneous set of objects (the rows of the input table), each object consisting of a fixed set of typed values taken in a finite domain. The names and types of these values define the attributes of the table; together the attributes define the table's schema. A table instance is set of objects that conforms to a particular schema. We could consider other input data structures, such as an XML document. The properties of our model would remain unchanged in essence, but would be much harder to formalize. In the following, we consider that a data table is accessed by means of a random access cursor: Row(integer) sets the current row being read, $attribute returns the current row value for the given attribute and Length returns the table instance's length.

### 3.2 Graphic language

The output of a visualization algorithm is an image. This image is created on computers by means of a graphic language, such as OpenGL or Java2D. Modern graphic languages consist of 2 types of instructions, whose parameters must be instantiated for each primitive through the result of a computation that depends (often) on the input data

**Geometric primitives**. Instructions such as fillRectangle(x,y,w,h), drawString(s,x,y)... These instructions define shapes (a priori) perceived by the user.

**Graphic attributes**. Instructions such as setColor(r,g,b), setFont(a), setPattern(p)… define graphic attributes which are interpreted in the context of drawing a geometric primitive. Since at most one graphic attribute of each kind is associated with each geometric primitive, counting geometric primitives alone is sufficient to estimate the output size of a visualization algorithm.

Graphic languages also have *configuration* and *transform* instructions. We will not consider these as the configuration instructions are usually called a fixed number of times at the beginning of the rendering pass, while transform instructions can be modeled as programming language operations.

### 3.3 Visualizations and Representations

A *visualization* is a function that maps a table instance onto a sentence of the graphic language, which will be called the *representation* of the instance. A visualization may depend on a particular schema, but we will not consider this potential compatibility problem. Computable visualizations are therefore regular programs, which distinguish themselves in that they contain data access instructions (input) and graphic language instructions (output). Example 1, below, is a trivial example of a visualization.

```
int i=0;
setColor("black");
for(int c=0;c<Length; c++) {
    Row(c);
    drawString($name,0,i);
    i+=20;
}
```

Example 1: a visualization written as an imperative program displaying the name of each object one below the other.

The representation of the input data (john, mary, tom) by the visualization given in example 1 is the sequence:
```
drawString("john", 0, 0);
drawString("mary", 0, 20);
drawString("tom", 0, 40);
```

A table representation does not fully convey how the resulting image will be perceived. Still, analyzing the visualization algorithm, and in particular, the dependencies between input attributes and output primitives, can yield exploitable observations on what information may be or may not be available to the user of a specific visualization.

## 4 USING LINEAR-STATE DATAFLOWS

Before presenting our model formally, a tour of its main features should provide an intuitive grasp of its properties. As an example, we use some statistics on American cities, containing values such as its name, State, population, crime rate, housing cost, climate, latitude and longitude. A visualization is defined as a declarative sequence of relationships between data variables (input) and image variables (output), typical of dataflow architectures.

### 4.1 Graphic primitives and graphic attributes

For each graphic primitive defined in the graphic language, our dataflow defines a language keyword. Graphic primitives can be assigned to each input object. For instance, the fillEllipse(x,y,w,h) primitive will have x, y, w and h parameters, which must all be defined by an expression.

The very simple visualization below (Example 2) assigns a filled ellipse to each row in the data set, whose coordinates are proportional to the latitude and longitude values held in the rows, while its width and height are of constant size. The iteration through all rows in the table is implicit, and the values held in the rows are implicitly mapped onto an interval from 0 to 1, which denotes the minimum and maximum positions in the drawing area. This mapping can be modified, for instance to handle logarithmic distributions, using a specific subparameter.

```
Visualization {
  FillEllipse {
      X = $Longitude;
      Y = $Latitude;
      Width = .04;
 // 4% of available width
      Height = .04;
    }
}
```

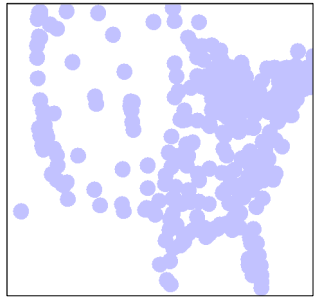

Example 2: Declarative program and resulting image of a 2D graph representing the latitude and longitude of each city.

```
Visualization {
  FillEllipse {
   // …other parameters…
    Paint {
      hue = .75;
      saturation = .5;
      value =
  $Population > 1M ?
      1 // blue
      : 0 // black
    }
  }
}
```

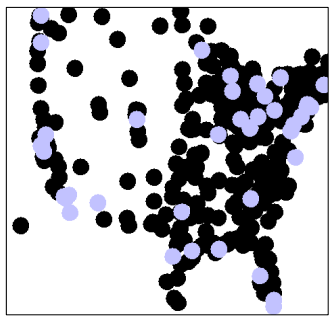

Example 3: cities above 1M inhabitants are drawn in blue.

Just as graphic primitives are assigned to data objects, graphic attributes can be associated with each primitive. Example 3 specifies a palette to highlight the cities above 1 million inhabitants.

Further convenience settings let the user specify that X and Y positions should be centered rather than left justified, and add some scales and grids to produce an acceptable 2d plot:

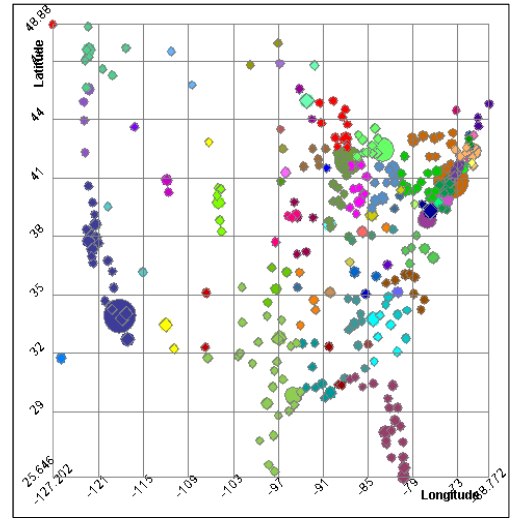

It is possible to add more graphic primitives defined within the context of another primitive. The embedded primitives are then defined in a local coordinate system, as shown by example 4.

```
FillRectangle {
 X=$Longitude;
 ...
 FillRectangle {
    X=0; Y=0;
    Width=0.5;
    Height=$Crime
 }
 FillRectangle {
    X=0.5; Y=0;
    Width=0.5;
    Height=$Climate
 }
}
```

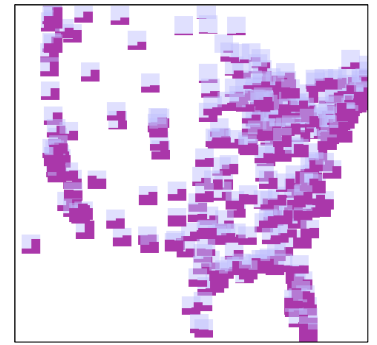

Example 4: two rectangles drawn inside the main rectangle of each city create a plot of small 2-valued bar charts. The rectangles have a height proportional to some chosen attributes.

So far, dataflow users should not have been surprised: many systems allow performing such kinds of mappings. Now, often in visualization, we want to place objects side by side rather than according to an attribute of the table. While this is very easily done with a regular imperative program (see Example 1 for instance), it requires some state handling that violates the regular dataflow paradigm.

One originality of our dataflow comes from the notion of local variables, which insert themselves in the dataflow in a way similar to the coroutine mechanism of the CLU programming language [19].

### 4.2 Local variables

Local variables represent the state information that may be needed in a visualization. Local variables are defined with a name, an initialization expression, which is evaluated once before each loop over the table rows, and an iteration expression, which performs any sort of state change in between rows. In example 5, we place the cities in the database next to each other to fill the whole display area, and show a histogram for each row:

```
Sort = $Population;
Variable {
  i= { init = 0;
       iter = i + 1/Length
     }
}
FillRectangle {
  X = i; Y=0; Height=1; // full height
  Width = 1/Length;
  FillRectangle {
    X=0; Y=0;
    Width=1;
    Height=$Population/3
  }
  FillRectangle {
    X=0; Y=1/3;
    Width=1;
    Height=$Climate/3
  }
  FillRectangle {
    X=0; Y=2/3;
    Width=1;
    Height=$Crime/3
  }
}
```

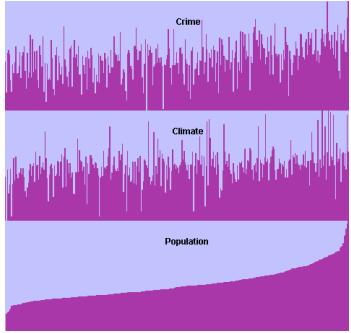

Example 5: parallel histograms

The above program associates a filled rectangle, of height 1 (full height) and y coordinate 0, to each row. The x coordinate of each rectangle is determined by a variable `i`, which accumulates the sum of the widths of the rectangles already drawn. The width of each rectangle is inversely proportional to the number of objects held in the table. Adding code-bar-like children graphic primitives for each record results in a "parallel histograms" visualization.

### 4.3 Accumulators

Accumulators specify passes through the data table (or part of it) that do not generate output. They are defined like local variables with an initial expression and an iteration expression, and are often used to compute values on a subset of the data table. To illustrate its use, suppose that instead of a regular "parallel histograms", we want the width of each rectangle to be proportional to the population held in each record (to put large cities forward, for instance). To achieve this, we need to specify that the width is no longer `1/Length`, but `$Population/Sum($Population)`. This results in an "adjustable width parallel histograms", which as far as we know is a new type of visualization, emphasizing a ratio attribute (Population) in a "parallel histograms" visualization (Example 6).

```
Accumulator {
  Sum= {
    init=0;
    iter=Sum+$Population
  }
}
Variable {
  i = {
    init=0;
    iter =
       i+$Population/
             Sum
  }
}
```

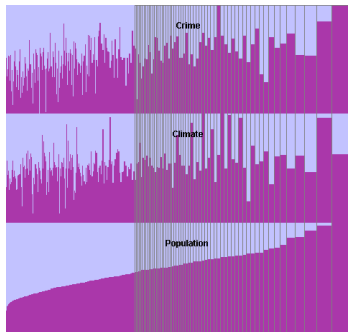

```
FillRectangle {
  X=i; Y=0; Height=1
  Width=$Population/Sum
  …
}
```
Example 6: "adjusted-width parallel histograms"

Accumulators and variables can be used to define many types of layouts, such as grids, spirals or squarified treemaps. However, they do not provide means to perform non-sequential searches in the data set, for instance to implement heuristic-based positioning of the primitives. This limitation is intentional: it keeps the model declarative and simple, but leaves it expressive enough to reach a precise class of useful visualization algorithms, as will be seen later.

### 4.4 Partitions and subvisualizations

Partitioning defines groups of objects that share some common visualization characteristics. It corresponds to conditional branches (if()/else or rather switch()/case instructions) in the inner loop of a visualization program.

For instance, assume that instead of visualizing the cities, we were more interested by the states to which they belong (Example 7).

```
Partition = $State {
  Accumulator {
  // average longitude and latitude
  // for each state/city
    AvgLong= {
      init=0;
      iter = AvgLong+$Longitude;
      end = AvgLong/recordCount;
    }
    AvgLat= {
      // idem…
    }
  }
  FillRectangle {
     X=AvgLong-0.05;
     Y=AvgLat-0.05;
     Width = 0.1;
     Height = 0.1
  }
}
```

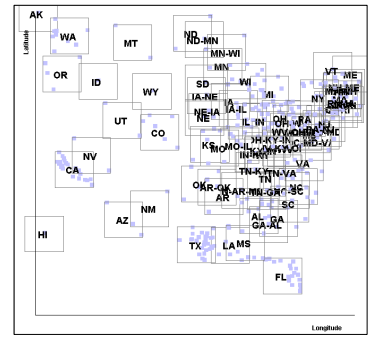

Example 7: 2D plot of states obtained by grouping cities belonging to the same state together and producing the same visualization as Example 2.

A few predefined accumulators are introduced to simplify notation: `childCount` is set to the number of partition elements, `recordCount` to the cardinality of a partition element and `depth` to the current depth of the partition. It must be observed that, by default, the *visualization definition is reapplied recursively to all partition elements*. Hence, without further specification, a 2D plot of the cities belonging to each state is produced inside the rectangle delimiting each state. This is particularly useful for defining recursive visualizations such as treemaps.

For more common representations, we can define subvisualizations. The hierarchy of partitions defines for each node (row or group of rows) a unique path. This path is used in the program to define subvisualizations specific to each partition element. A subvisualization therefore corresponds to a subroutine in a visualization program that is being called conditionally for rows of the table that match some common criteria. Example 8 lays out the states on a grid, showing a 2D plot for each city, except for California (CA).

Defining the grid requires 2 accumulators to find the number of rows and columns given the number of children in the partition, and a variable `i` to iterate for each State through the rows and columns.

```
Partition = $State {
  Accumulator {
      Rows=sqrt(childCount);
      Columns=floor(sqrt(childCount-1))+1
  }
  LocalVariable {
      i={init=0; iter=i+1}
  }
  FillRectangle {
      // place states on a grid
      X=(i%Columns)/Columns;
      Y=floor(i/Columns)/Rows;
      Width=1/Columns;
      Height=1/Rows;
  // define per partition element subvisualizations
      Children {
        * { // * is a wildcard character
              // meaning all children:
              // we define a 2D plot
           X=$HousingCost;
           Y=$Climate;
            …
        }
        CA { // California will have a different
             // representation
           …Paint=black…
        }
      }
   }
}
```

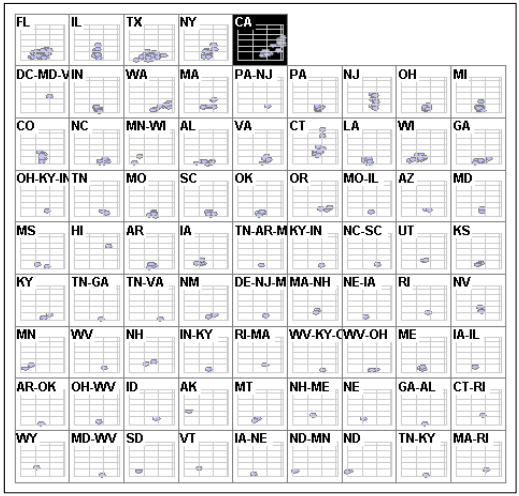

Example 8: a "Grid of 2D graphs", enables comparing variations in trends between states, here ploting Housing cost vs. Climate for each state and treating california (CA) with a subvisualization that changes its background color.

### 4.5 Recursive partitioning
As seen in example 7, Partitioning is recursive by default. We can specify a partitioning expression whose value depends on the current depth level, using the `depth` variable,

For instance, specifying the expression {$State, $County, $City}[depth] as a partitioner will first group the objects by State, then by county, and finally by city (assuming we have a set of neighborhoods as input).

Consider a data table representing files in a file system. A "Path" attribute contains the "/"-separated path name of each file, such as "/usr/lib/sendmail/sendmail.cf". A treemap (Example 9) is obtained by partitioning the input table with a formula that splits the Path attribute into an list, using "/" as a separator. A "Horizontal" variable enables alternating vertical and horizontal packing.

```
Partition = split($Path,"/")[depth] {
    Accumulator {
       Sum= { init=0; iter=Sum+$FileSize;}
       Horizontal=depth%2;
    }
    LocalVariable {
       Position= {
           init=0; iter=Position+$FileSize/Sum
       }
    }
    FillRectangle {
       X=Horizontal?0:Position;
       Y=Horizontal?Position:0;
       Width= Horizontal?1:$FileSize/Sum;
       Height=Horizontal?$FileSize/Sum:1;
    }
}
```

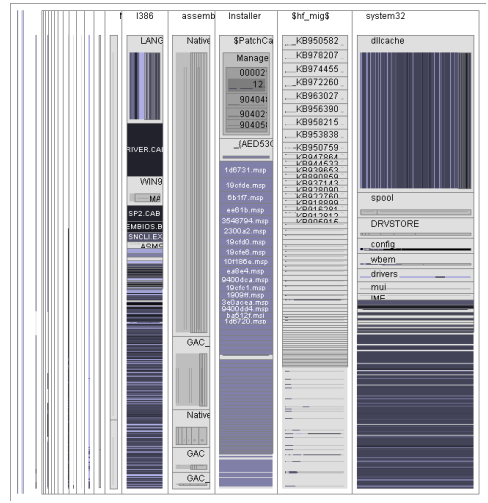

Example 9: file hierarchy of 25000 files in a treemap

Squarified treemaps are obtained by constructing a partitioner that alternates at each depth level between "Path" elements and a local variable that keeps track of the aspect ratio of the current subset of the rows being partitioned, yielding a new partition element each time the heuristic has reached a local optimum.

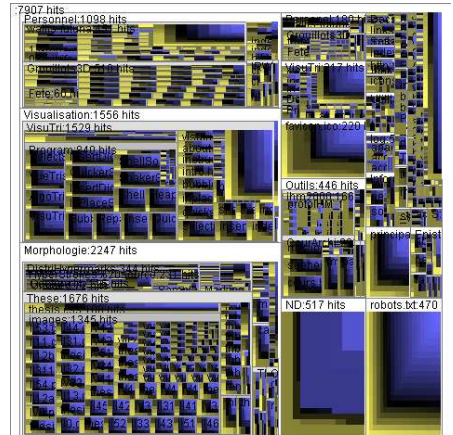

Example 10: web log of 10000 hits in a (squarified) treemap.

## 4.7 Ordering and Filtering

In Example 5, we've seen a simple usage of the Sort operator. This operator is in fact subtler, as it must provide the ability to also order partition elements relative to each other or to individual rows. To achieve this, Sort accepts the specification of inner accumulators, which return a single value for each partition element. Those values are used in the comparison method. Predefined accumulators such as Sum or Average are provided to handle the most common uses.

In our implementation, we use a comparison-based sorting method, of complexity O(n log(n)). To fully comply with data-linear visualizations as introduced thereafter, we should instead provide an Order operator that can return arbitrary permutations of the input indices obtained by performing at most a bounded number of passes through the input. This operator functions like an Accumulator whose result is an array of indices, and can embed additional pre-computations in the form of sub-partitions, sub-accumulators and local variables. Order allows implementing sort algorithms of linear complexity, such as radix or bucket sorts of fixed maximal depth. This maximal depth would be equal to the logarithm of the cardinality of the domain of the table attributes. In practice, our Sort operator combined with a Filter operator that removes row indices is just more convenient.

## 4.8 Other features and representations

To allow our implementation to fully respect its corresponding computational model, we also include a RepeatGeometry operator that provides a mechanism for assigning an unbounded number of geometric primitives to a row or group of rows. To avoid entering complex expressions to produce often needed functionality, our language allows the definition of margins, scales, grids, labels...; Macros enable setting variables and accumulators in one step for common display schemes, such as placing the objects next to each other horizontally or vertically. Furthermore, as we have mentioned earlier, we offer features to align the data domains to the geometrical domains: when an attribute is provided as input, the user can specify on which interval (from the root or from a local group) and according to which function (linear, adjusted, spread…) must the attribute be projected onto the geometry.

Many other types of visualizations can be obtained by interactively specifying program parameters: parallel coordinates, 2D graphs of histograms, matrices of histograms… Furthermore, the user can choose to represent different portions of the data set with specifically tuned visualizations. The following visualization synthesizes the execution of a search on the 39 queens problem, a classic problem of constraints programming. The visualization superposes the search tree (shown as a treemap and reinforced by the varying gray levels) onto 10 variables summarizing the state of the constraint solver at each decision step of the search tree.

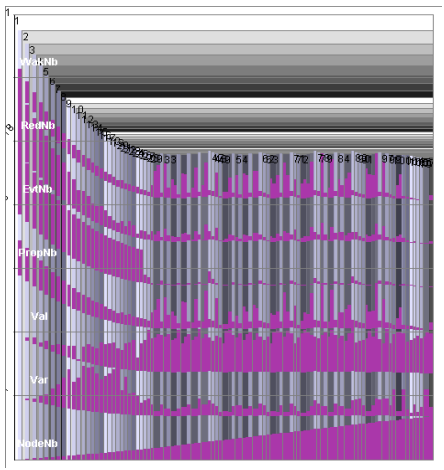

Example 11: parallel histograms structured as a (non-alternating) treemap. Defining this visualization requires 6 interactive settings.

To experts trained on the optimization of constraint propagation, rich views such as this one can help quickly identify pathological cases and help the understanding of domains as complex as constraints-based programming..

Parallel coordinates [18] are produced by using a "Polyline" graphic primitive and assigning its y coordinate to data attributes, while its x coordinates are evenly spaced:

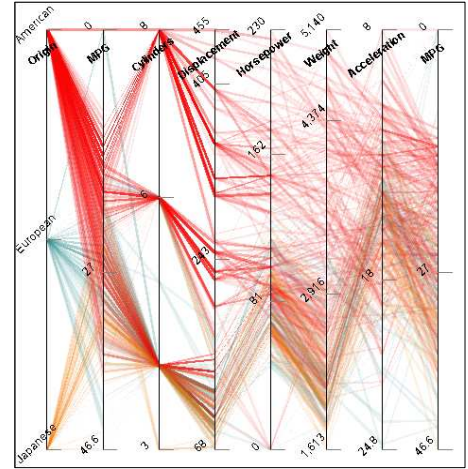

Example 12: parallel coordinates

## 5 LINEAR STATE DATAFLOWS

The declarative visualization language we have introduced in the previous section models a dataflow consisting of four major operators: partitioning, sort/order, graphic primitive assignment (including RepeatGeometry) and decoration assignment. State handling operators (variables and accumulators) can be attached and scoped to each node of this dataflow, enabling some of the operations ordinary handled by imperative programs, namely performing single linear passes over all rows of the table. Hence, we call our model linear-state dataflows, summarized in figure 2.

All the parameters of the dataflow (partition method, sort criteria, types and parameters of assigned primitives…) are expressions of the programming language that do not call Row() nor perform output instructions. These expressions can still access the local variables, accumulator values, and current row values (through $att calls).

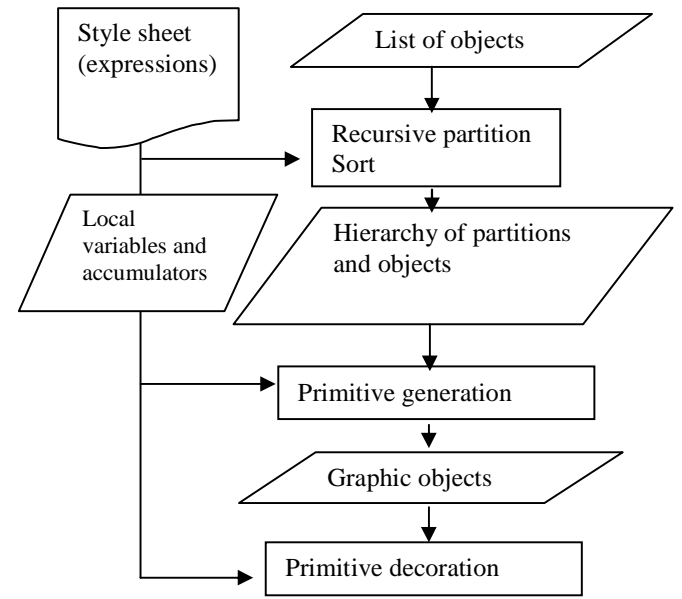

Figure 2: Linear state-dataflows

As we have seen in the previous section, numerous types of visualizations can be produced with this augmented dataflow model. This model allows assigning graphic primitives to one row or to a group of rows, allows some kinds of loops, has state handling and conditional representation capability. Yet, many algorithms, such as spring embedded graph layout would require additional operators to be representable with the dataflow. Namely, to achieve this kind of primitive placement, one would require an accumulator that allows performing an unbounded number of passes through the data set to determine the positions of the nodes.

Given that the linear-state dataflows provide the basic building blocks of regular imperative programs, we wish to characterize more formally the class of programs that can be represented with this model, much like relational calculus is tied to relational algebra in database querying. In the following, we show that linear-state dataflows are equivalent to the class of programs that perform at most a fixed, predetermined number of passes through the dataset. Furthermore, we will see that linear-state dataflows are a canonical model: they define an equivalence class over data-linear visualization algorithms which should provide a starting point to compare and classify visualization algorithms properties in a space of limited dimensionality.

In the next 3 sections, we summarize the proof of this equivalence. We first define data-linear visualizations. Then we show a lemma introducing a canonical representation of all visualization algorithms. Then we simplify this canonical representation for data-linear visualizations. Finally, we show how combinations of linear-state dataflows operators can be used to model all the components of a data-linear visualization in its canonical form. The reverse direction is trivial, as linear-state dataflows do not allow expressing more than a fixed number of passes through the input dataset.

## 6 INPUT (OR DATA) COMPLEXITY

There are several ways to define a computational complexity for visualizations. Algorithmic complexity relies on the number of instructions that are executed when the program is run. Output (or rendering) complexity relies on the number of graphic primitives produced as output. Input (or data) complexity, on which we focus, relies on the number of times each input data object is accessed. We focus on worst-case complexity, as it is easier to handle than average-case complexity. We define the (input/data, worst-case) complexity of a visualization as the function C that maps the table length (L) to the maximum number of times `Row()` is called during the execution of the program, for all table instances of length L.

For example, the complexity of the visualization in example 1 is the identity function: for all n, C(n)=n. For input complexity to be properly defined, we must assume that the input data size is arbitrarily large, while the available memory (held in the state array described below) is bounded by a sufficiently large value. Otherwise, all visualizations could be made input-linear by loading the full table in an array once and never calling `Row()` after that.

Note that the algorithmic complexity is always superior or equal to the input complexity: the program "`while(true) {}`" will loop indefinitely, but its input complexity is 0, as it never calls the `Row()` function. In most useful programs, though, input complexity is closely tied to algorithmic complexity, as exemplified in the extensive literature on searching and sorting.

We choose to investigate input complexity rather than pure algorithmic complexity for a reason: this measure provides a good estimate of how "hard" must the input be "worked over" to produce an image. Assuming the output image is not underutilizing the processing done, it provides an estimate of the amount of information extracted from the table and contained in the output image. Assessing how much of the contained information is actually perceived by users through controlled experimentation would provide a mean to define a "visualization efficiency measure" for a particular visualization.

### Data-Linear visualizations

A visualization is (worst-case) data-linear when there is a constant K such that, for any table instance, for all i<Length, Row(i) is called at most K times. Its complexity is bounded by K*Length, but this definition is more restrictive than saying that its complexity is a linear function of Length. [1]

Besides their having simple canonical forms, as will be seen later, data-linear visualizations are quite interesting by themselves:
- They are the smallest class of "interesting" visualizations: displaying exact information on the data set as a whole requires at least one pass through all rows.
- They are usually very fast and efficient, as one can bound the number of passes made on the data set.
- Many useful visualizations are data-linear: 2D graphs, histograms, but also most kinds of treemaps, space-filling visualizations, and recursive representations of trees are data-linear.

The typical useful visualizations that are not data-linear are sophisticated graph-drawing algorithms. The fast Fourier transform is also not data-linear, while short-time fast Fourier transforms (with a fixed window size) are. In general, visualizations that require repetitive searches through the data set to place graphic primitives according to values held in two or more non-consecutive rows (such as heuristic placement) are not data-linear.

## 7 CANONICAL REPRESENTATION OF VISUALIZATIONS

We now introduce a lemma showing that we can reorganize any visualization program as a (possibly infinite) sequence of single, ordered, passes through the dataset. Each of these passes will be interpretable as a single state-dataflow sequence taking the table as input, passing it through filtering, ordering and primitive generation operators to output a representation.

More formally, our lemma states that any visualization algorithm can be rewritten as a program following the canonical structure of Program 1 without changing its complexity by more than a linear factor.

```
Data state=initialization();
while(NeedMorePasses(state)) {
  Iterator a=OrderInput(state);
  PerPassInitialization(state);
  PerPassOutput(state);
  for (int i=0;i<a.size;i++) {
    Row(a[i]);
    PerRowOutput(state);
    PerRowIteration(state);
  }
  PerPassPostOutput(state);
  PerPassIteration(state);
}
```
1 pass through the data set = 1 linear-state dataflow operators sequence

Program 1: canonical representation of visualization algorithms

Wherein:
- The `OrderInput()` function returns an ordered set of row indices. Therefore, each row is accessed at most once in the following inner loop.
- All the functions cited (`NeedMorePasses`, `PerPassIteration`...), except `OrderInput`, do not call `Row()`.
- The calls to graphic language functions only happen in the `*Output()` calls, and those do not modify the state array.

---

[1] To make this clear, we can imagine a visualization that calls `Row()` sqrt(Length) times over sqrt(Length) rows: this visualization is not data-linear by our definition, even though its complexity is a linear function of the input size.

In this representation, visualizations are equivalent when, given identical input and state information:
- All functions have the same side effects on the state.
- `*Ouput()` functions produce identical output.
- `OrderInput()` return the same ordered set of indices.
- `NeedMorePasses()` functions return identical values.

Note that knowing whether or not two visualizations are equivalent is not decidable in the general case.

### 7.1 Canonical form of imperative programs

We now demonstrate this lemma using a Java-like pseudo-language. The proof works by executing the program with an interpreter that takes as input a visualization and a data table, interprets its instructions while memoizing its state changes, but without performing output. The interpreter proceeds until it has detected that a single pass has been performed; then it performs this pass again, but this time executing the corresponding output instructions. The interpreter loops until it has reached the last instruction of the input visualization. As this interpreter's control flow fits the canonical representation, any visualization can be put in canonical form.

More precisely, given a visualization A, the interpreter creates a derived version A* wherein calls to `Row(i)` are replaced by a macro `RowD(i)` and calls to `Draw(params)` are replaced by a macro `StoreDraw(params)`, defined thereafter. We also define a memoization data structure, called Iterator, that will retain the order in which the rows are accessed and the local context associated to each `Draw()` call:

```
class Iterator {
  sequence<rowindices> rows;
  associativetable<row,params> instructions;
}
```

A* is then used in Program 2 to execute the visualization:

```
constant { preIteration, postIteration};
global boolean needMorePasses=true;

Data state=initializeState();
while(needMorePasses) {
  Iterator a=OrderInput(state);
  // PerPassInitialization empty
  // PerPassOutput:
  a.interpretDrawInstructions(preIteration);
  for (int i=0;i<a.size;i++) {
    Row(a[i]);
    // PerRowOutput:
    instructions.get(a[i]).execute();
    // PerRowIteration empty
    // because draw params are memoized
  }
  // PerPassPostOutput
  a.interpretDrawInstructions(postIteration);
}

global Data storedState;
function OrderInput(state) {
   Iterator result=new Iterator();
   boolean visitedRows[Length];
   int currentRow=preIteration;
   if(storedState!=null)
       state=storedState;
   Interpret(A*);
   needMorePasses=false;
   return result;
}
```

Program 2: arbitrary visualization A put in canonical form.

The function `OrderInput` iterates through the rows, checking if the row has already been accessed before, and exits as soon as it has detected a return on a previous row. It performs memoization of the draw parameters and stores the interpreter's state appropriately to be able to restart where the program left.

The macros `RowD` and `StoreDraw` are defined in the context of `OrderInput`:

```
macro RowD(i) {
  if(visitedRows[i]==false) {
      Row(i);
      currentRow=i;
      result.rows.add(i);
      visitedRows[i]=true;
  } else {
      storedState=state;
      return result;
  }
}
macro StoreDraw(params) {
      result.instructions.add(currentRow, params);
}
```

A more constructive proof may be devised. It would use a meta-interpreter that captures the instructions leading to a particular draw call and inserts those instructions in the flow of `PerPassInitialization` and `PerRowIteration`. This would avoid having to rely on massive memoization and produce a program that conforms better to the original. We don't need such a detailed approach for handling data-linear visualization.

### 7.2 Canonical form of data-linear visualizations

The canonical form defined above is quite impractical to manage in the general case, as the number of outer loops increases roughly like the input complexity of the visualization. However, this representation can be quite simplified for data linear visualizations: it stems from their definition that their outer loop consists of a fixed, predetermined number of passes:

```
// variables and accumulators declarations
Data state=initialization();
for(int j=0;j<K;j++) {
    // accumulators, partitions and ordering
    Iterator a=OrderInputR[j](state);
    // convenience operators: scales, legend…
    PerPassOutput[j](state);
    // variables initialization
    PerPassInitialization[j](state);
    for (int i=0;i<a.size;i++) {
       Row(a[i]);
       // graphic primitives
       PerRowOutput[j](state);
       // variables iterations
       PerRowIteration[j](state);
    }
    // convenience operators: label, frame…
    PerPassPostOutput[j](state);
}
```

Program 3: canonical representation of data-linear visualizations and its correspondance with linear-state data-flow operators.

Furthermore, the `OrderInputR` functions are restricted to perform only a bounded number of iterations over the input (ruling out comparison-based sorting, which we introduce in our implementation only for convenience).

As a consequence, a data-linear visualization of complexity K can be canonically represented by an ordered set of at most K independent functions [`OrderInputR`, `PerPassOutput`, …] respecting the constraints set in our lemma.

## 8 TRANSFORMING THE CANONICAL FORM IN LINEAR-STATE DATAFLOWS

We will now describe how each pass in a data-linear visualization can be modeled with a single linear-state dataflow operator sequence. To this effect, we describe how each of the functions described in the canonical form can be produced by a combination of linear-state dataflow features. As a consequence we will have shown that any data-linear visualization can be modeled by an ordered set of linear-state dataflow operator sequences.

The `PerPass-` and `PerRow- Output` functions are trivially modeled by a sequence of *Graphic Primitive* operators, possibly embedded in a *RepeatGeometry* operator to provide unbounded output complexity. These may be placed in the context of *Filter* or *Children* nodes to allow conditional generation of drawing operations. The `PerPassInitialization` and `PerRowIteration` functions are, as evidently, modeled by *LocalVariables* operators.

The `OrderInputR` function can be modeled by a combination of *Order*, *Accumulator* and *Partition* nodes. The maximum number of passes through the dataset may depend on the row content, as exemplified by the treemap implementation (4.5). Yet, as we have mentioned in 3.1, the table attributes have a finite domain. In consequence, we can determine a fixed boundary on the number of passes through the data set that may be triggered by the content of one or more rows. Because *Partition* nodes can be recursively or selectively reapplied to subsets, any iteration order through the data set relying on the table content can be defined. Assuming the *Partition*, *Order* and *Children* operators use a Turing-complete language in their expressions, any (linearly-constructed) order of traversal can be defined for a particular iteration. As a result, any `OrderInputR` function performing a fixed number of passes through the dataset to yield a given ordering of traversal can be modeled with linear-state dataflows. The proof of this modelisation relies on a technique similar to the demonstration of the lemma on canonical form. However, it requires a precise delimitation of the operators' semantics, which goes beyond the scope of this article.

With the *Sort* operator, our implementation is a little bit more powerful than pure data-linear visualizations. However, replacing it with the less convenient *Order* operator, there is no way in linear-state dataflows to express an algorithm that runs over the dataset for an unbounded number of passes. Therefore, visualizations defined with a linear-state dataflow and data-linear visualizations are isomorphic through a transform of constant algorithmic complexity.

The additional features of the *Partition*, *Sort* and *Filter* operators are provided to reproduce common structuring features found in programming languages. For instance, the local definition of accumulators and variables inside a *Partition* operator is a mean to reproduce the local context (stack frames) associated to imperative language functions. While linear-state dataflows are certainly not the only way to conveniently model data-linear visualizations, our walkthrough should provide enough of an understanding of their expressiveness and conciseness.

## 9 CONCLUSION

The linear-state dataflow model has been implemented (modulo a few details) in an information visualization framework [3, 4, 17]. Among other features, it provides an editor and a real-time viewer that enable interactive construction of data-linear visualizations. Note that a macro mechanism enables creating the visualizations shown as examples directly, without requiring knowledge of the underlying model.

Besides being expressive, our model is also very efficient and enables automatic cache handling to trade-off speed for memory usage: all or part of the expression's evaluations can be cached in the dataflow nodes. This enables very efficient rendering: the refresh rate will stay interactive or close to interactive with data sets of 1 million objects or more.

The important aspect to retain from our model, in our opinion, is that it defines a canonical representation of the design space of data-linear visualizations, which are a large and useful class of visualizations. This offers important perspectives, both for the advancement of theory and the practice of information visualization:

From a theory point of view, we have provided a declarative model that describes canonically the full design space of an important set of visualization algorithms. This gives us a basis to better understand the characteristics of various representations, define metrics on this space, and analyze the differences between this space and a still-to-be-defined corresponding perceptual space of visualizations ([5], [32], [36]).

From a practical point of view, our model lets us define a style sheet-like interface and an interactive visualization editor, which are powerful, flexible, and particularly efficient. Because our model only defines a fixed number of parameters and a fixed interconnection between each node of the dataflow, it is expected that this model should be easier to use than Turing-complete programming models.

The model we propose is still targeted at expert users, but our editor provides a macro mechanism and ready-to-use visualizations, enabling lay users to obtain variations of common visualizations with very few settings. Finally, we hope to use the linear-state dataflow model to define quantitative metrics of the information displayed by various visualization algorithms.

## ACKNOWLEDGMENTS

F. van Ham, B. Haible and G. Sander contributed through major comments on the model. I thank the rest of the Discovery team: R. Dupuy, S. Haas and the technical writers. Also, some Discovery users: C. Lepape, P. Deransart, L. Langevine, and R. Dumeur, have provided many usability improvement suggestions.